\documentclass[preprint,aps,12pt,showpacs,nofootinbib,tightenlines,amsmath,amssymb]{revtex4}
\usepackage{amsmath}
\usepackage{graphicx}
\usepackage{amssymb}
\usepackage{color}
\newcommand{\VEV}[1]{\left\langle #1\right\rangle}

\newcommand{\p}{\partial}

\newcommand{\MeV}{\;\text{MeV}}

\begin{document}
\def\intdk{\int\frac{d^4k}{(2\pi)^4}}
\def\sla{\hspace{-0.17cm}\slash}
\hfill

%\begin{titlepage}
\title{Thermal Spectral Function and Deconfinement Temperature in Bulk Holographic AdS/QCD with Back Reaction of Bulk Vacuum}

\author{Ling-Xiao Cui}\email{clxyx@itp.ac.cn}
\author{Zhen Fang}\email{fangzhen@itp.ac.cn}
\author{Yue-Liang Wu}\email{ylwu@itp.ac.cn}

\affiliation{State Key Laboratory of Theoretical Physics(SKLTP)\\
Kavli Institute for Theoretical Physics China (KITPC) \\
Institute of Theoretical Physics, UCAS \\
Chinese Academy of Sciences, Beijing, 100190, China}

\date{\today}

\begin{abstract}
Based on the IR-improved bulk holographic AdS/QCD model which provides a consistent prediction for the mass spectra of resonance scalar, pseudoscalar, vector and axial vector mesons,  we investigate its finite temperature behavior. By analyzing the spectral function of mesons and fitting it with a Breit-Wigner form, we perform an analysis for the critical temperature of mesons.  The back-reaction effects of bulk vacuum  are considered, the thermal mass spectral function of resonance mesons is calculated based on the back-reaction improved action. A reasonable melting temperature is found to be $T_c \simeq 150 \pm 7$ MeV, which is consistent with the recent results from lattice QCD simulations.
\end{abstract}
\pacs{12.38.Aw,12.38.Lg,11.15.Tk,11.10.Wx}

\maketitle

\section{Introduction}
\label{Chap:Intro}

The property of asymptotic freedom of quantum chromodynamics (QCD)\cite{AF} and the treatment of non-perturbative QCD have led to the QCD string approach, which has eventually initiated the motivation of string theory. with the development of string theory, it further motivated the advent of the AdS/CFT conjecture \cite{Maldacena:1997re,Gubser:1998bc,Witten:1998qj,Polchinski:2001tt}, which provides an alternative tool to access the gloomy non-perturbative region of QCD, that is so-called holographic QCD or AdS/QCD model based on the AdS/CFT. These models are not perfect with some problems in its deep root of AdS/CFT as  QCD is not a conformal field theory at low energy.  There are different holographic QCD models due to different realizations and objectives. It has mainly been divided into two classes, namely top-down model and bottom-up model. The top-down models are directly constructed from string theory, the popular ones like D3/D7, D4/D6 and D4/D8 model \cite{top down 1,top down 2,top down 3}. while bottom-up models such as hard-wall model \cite{hard wall} and soft-wall model \cite{soft wall} are constructed according to properties of QCD itself from which the corresponding bulk gravity is determined. In the hard-wall model, a sharp cutoff of the fifth dimension which corresponds to the inverse of the QCD scale $\Lambda$ is given to realize the QCD confinement. It contains chiral symmetry breaking but fails to give a correct Regge behaviour for the mass spectra of hadrons. To remedy this problem, in the soft-wall model, a dilaton term is put into the bulk action to replace the sharp IR cutoff of the hard-wall model. However, the resulting model cannot realize chiral symmetry breaking phenomenon consistently. Several models have been constructed to incorporate these QCD behaviors\cite{quartic term,Sui:2009xe, Sui:2010ay,Cui:2013xva,Li:2012ay,z bulk mass2,z bulk mass3,BT1,BT2,BT3}, these models have made numerical predictions for the mass spectra of light mesons, such as scalar, pseudoscalar, vector and axial-vector mesons. Especially, in the recent paper\cite{Cui:2013xva}, we have constructed an alternative model in which the metric remains conformal invariance and satisfying Einstein equation, while the bulk mass and bulk coupling of the quartic scalar interaction have a bulk coordinate z-dependence, so that the ultraviolet (UV) behavior of the model corresponds to AdS/CFT, while the infrared (IR) behavior is required from low energy QCD features which are compatible with the leading chiral dynamic model of spontaneous chiral symmetry breaking\cite{Nambu:1960xd,DW}. As a consequence, we have arrived at a more consistent model with better predictions for the mass spectra of both ground and resonance states of scalar, pseudoscalar, vector and axial-vector mesons.

The finite temperature effects of holographic QCD has attracted lots of attention.  The finite temperature effects in hard-wall AdS/QCD were studied in \cite{hardwall_T}. In \cite{Fujita:2009wc,Fujita:2009ca,Miranda:2009uw,Colangelo:2009ra}, the thermal spectrum of glueballs or mesons in the soft-wall AdS/QCD model was investigated. In \cite{Grigoryan:2010pj}, a soft-wall model for charmonium was built. The deconfinement temperature of soft-wall AdS/QCD models was calculated in \cite{Herzog:2006ra} and found to be $T_c\approx191\MeV$. In Ref.\cite{Scalar glueball,glueball_1,glueball_2}, the scalar glueball and light mesons spectral have been analyzed in the soft-wall AdS/QCD model and the critical temperature at which the meson states dissociation was found to be about $40-60\MeV$. Such a low temperature is far from the deconfinement transition. It indicates that the meson states dissociation occurs in the confined QCD phase and it is inconsistent with the real QCD. To remedy this problem, we have investigated in \cite{Cui:2011ag,Cui:2013zha} the finite temperature effects for the metric IR-improved soft wall AdS/QCD models \cite{Sui:2009xe}. The critical temperature of meson dissociation was found to be around $200\MeV$. Where the metric is modified at IR region, so the Hawking temperature of the black hole is not exactly defined as it dose not satisfy Einstein equation. Thus it is interesting to analyze the critical temperature of the bulk holographic AdS/QCD model built recently in\cite{Cui:2013xva}, where the model incorporates both chiral symmetry breaking and linear confinement with the better predictions on the mass spectra of meson states.

The paper is organized as follows:  In Sec.\ref{Chap:Model}, by briefly reviewing the IR-improved bulk holographic AdS/QCD model constructed recently in\cite{Cui:2013xva} , we extent it to an action with finite temperature. In Sec.\ref{Chap:thermal spectral}, we analyze the thermal spectral function and carry out calculations for the meson thermal mass spectra. The corresponding melting temperature is obtained. In Sec.\ref{Chap:back-reaction}, the back-reaction effort of bulk vacuum is considered to yield an improved metric of background gravity,  the thermal mass spectra are investigated in detail based on the back-reaction improved action. A reasonable melting temperature is obtained. Our conclusions and remarks are presented in the final section.

\section{IR-Improved Bulk Holographic AdS/QCD Model with Finite Temperature}
\label{Chap:Model}

In this section, we will investigate the finite temperature behavior of the IR-improved bulk holographic AdS/QCD model\cite{Cui:2013xva}. Here the AdS black hole is chosen as the background to describe temperature in boundary theory,
\begin{equation} 
ds^{2}=\frac{R^{2}}{z^{2}}\left(f(z)dt^{2}-d\vec{x}^{2}-\frac{dz^{2}}{f(z)}\right),\label{metric}
\end{equation}
with
\begin{equation}
 f\left(z\right)=1-\frac{z^4}{z_h^4},
\label{AdS black hole}
\end{equation}
where $z_h$ is the location of the outer horizon of the black-hole. We will set the AdS radius as unity in this paper for the boundary theories. The Hawking temperature which corresponds to the temperature in boundary theory is defined as follow:
\begin{eqnarray}\label{hawking T}
T_{H}=\frac{1}{4\pi}\left|\frac{df}{dz}\right|_{z\rightarrow z_{h}}=\frac{1}{\pi z_{h}}
\end{eqnarray}

The action with finite temperature is based on the IR-improved bulk holographic AdS/QCD model\cite{Cui:2013xva}.
\begin{equation}
S=\int d^{5}x\,\sqrt{g}e^{-\Phi(z)}\,{\rm {Tr}}\left[|DX|^{2}-m_{X}^{2} |X|^{2}-\lambda_X |X|^{4}-\frac{1}{4g_{5}^2}(F_{L}^2+F_{R}^2)\right],\label{action}
\end{equation}
with $D^MX=\p^MX-i A_L^MX+i X A_R^M$, $A_{L,R}^M=A_{L,R}^{M~a}t^a$ and ${\rm{Tr}}[t^at^b]=\delta^{ab}/2$. The gauge coupling $g_5$ is fixed to be $g_5^2 = 12\pi^2/N_c$ with $N_c$ the color number \cite{hard wall}. The complex bulk field $X$ will be written into the  scalar and pseudoscalar mesons, the combination of chiral gauge fields $A_L$ and $A_R$  will be identified to the vector and axial-vector mesons. The dilaton field, the bulk scalar mass and quartic interaction coupling  have been shown to be reasonable to take the following IR-modified forms\cite{Cui:2013xva}:
\begin{eqnarray}
 \Phi(z) &=& \mu_g^2z^2-\frac{\lambda_g^4\mu_g^4z^4}{(1+\mu_g^2z^2)^3}. \\
  m_{X}^2(z) &=& -3-\frac{\lambda_1^2\mu_g^2z^2+\lambda_2^4\mu_g^4z^4}{1+\mu_g^2z^2} + \tilde{m}^2_X(z) \\
  \lambda_X(z) &=& \frac{\mu _g^2 z^2}{1+\mu _g^2 z^2}\lambda\
\end{eqnarray}
with $\lambda_1=\lambda_2 = \sqrt{2}$. The expectation value of bulk scalar field $X$ has a z-dependent form for two flavor case:
\begin{equation}\label{VEV}
\VEV{X}=\frac{1}{2}v(z)\left(
                         \begin{array}{cc}
                           1 & 0 \\
                           0 & 1 \\
                         \end{array}
                       \right)
\end{equation}
The bulk vecuum expectation value (bVEV) $v(z)$ with proper IR and UV boundary conditions has been taken the following simple form \cite{Cui:2013xva}:
\begin{equation}\label{bVEV}
    v(z)=\frac{A z+B z^3}{1+C z^2}.
\end{equation}
with
\begin{equation}\label{ABC}
     A=m_q\zeta,\quad B=\frac{\sigma}{\zeta}+m_q\zeta C,\quad C=\mu_c^2/\zeta
\end{equation}
and the coupling constant $\lambda$ is related to the vacuum expectation value via the equation of motion
\begin{eqnarray}
 v_q\equiv \sqrt{\frac{(2\mu_g)^2}{\lambda} } =  \frac{B}{C} =  \frac{\sigma }{\mu_c^2} + m_q  \zeta 
\end{eqnarray}

The involving five parameters have been fixed from the low energy parameters of mesons\cite{Cui:2013xva} and their values are represented in Table \ref{Table:parameter} .
\begin{table}[!h]
\begin{center}
\begin{tabular}{ccccc}
\hline\hline
 $\lambda_g$ &$m_q$(MeV) & $\sigma^{\frac{1}{3}}$(MeV) & $\mu_g$(MeV)  & $\mu_c$(MeV) \\
\hline
1.7 & 3.52 & 290 & 473 & 375   \\
\hline\hline
\end{tabular}
\caption{The values of five parameters}
\label{Table:parameter}
\end{center}
\end{table}

%20140309

\section{Thermal Spectral Function}
\label{Chap:thermal spectral}

The bulk scalar field can be decomposed as $X(x,z)\equiv(v(z)/2+S(x,z))e^{2i\pi(x,z)}$,
where $S(x,z)$ is the scalar meson field and $\pi(x,z)=\pi^a(x,z)t^a$ the pseudo-scalar field.
The chiral gauge fields can be combined into vector field $V^a_M$ and axial-vector field $A^a_M$ as
\begin{equation}\label{combine}
    V^a_M\equiv\frac{1}{2}(A^a_{L,M}+A^a_{R,M})\qquad \textrm{and} \qquad A^a_M\equiv\frac{1}{2}(A^a_{L,M}-A^a_{R,M}).
\end{equation}
The equations of motion for the meson fields are given as follows in momentum space by performing the Fourier transformation.
\begin{eqnarray}
\label{eq:eomV}
\textrm{V}&:&
V_x''(z)+\left(\frac{a'(z)}{a(z)}+\frac{f'(z)}{f(z)}-\Phi'(z)\right)V_x'(z)+\frac{\omega^2 V_x(z)}{f^2(z)}=0,\\
\label{eq:eomAV}
\textrm{AV}&:&
A_x''(z)+\left(\frac{a'(z)}{a(z)}+\frac{f'(z)}{f(z)}-\Phi'(z)\right)A_x'(z)+\frac{\omega^2 A_x(z)}{f^2(z)}+g_5^2\frac{v^2(z)}{z^2f(z)}A_x(z)=0\\
\textrm{S}&:&
S''(z)+S'(z) \left(\frac{3
   a'(z)}{a(z)}+\frac{f'(z)}{f(z)}-\Phi '(z)\right)\nonumber
   \\ && \qquad\qquad\qquad +S(z)\left(\frac{\omega^2}{f(z)^2}-\frac{a(z)^2 m_X^2(z)}{f(z)}-\frac{3\lambda_X(z)a(z)^2 v(z)^2}{2f(z)}\right)=0,\\
%=====
\label{eq:eomPST}
\textrm{PS}&:&
\pi''(z)+\pi'(z) \left(\frac{3 a'(z)}{a(z)}+\frac{f'(z)}{f(z)}+\frac{2
   v'(z)}{v(z)}-\Phi '(z)\right)
 +\frac{\omega ^2 \pi(z)}{f(z)^2}=0,
\end{eqnarray}
 Note that with the temperature increaseing, the horizon of black hole $z_{h}$ moves from infinity to boundary side.  Thus the solutions of equations of motion will drop into black hole before they vanish, so that one cannot use the method of finding eigenmodes. Alternatively,  we shall consider spectral function which is the imaginary part of the retarded Green's function. In the above equations, we have put three-momentum to zero:$\overrightarrow{p}=0$, which leads the retarded Green's function to be  simplified as: $G^R_{tt}=0$, $G^R_{xx}=G^R_{yy}=G^R_{zz}\equiv G^R(\omega)$. For equation of pseudo-scalar field, we have ignored the mixing between axial-vector field and pseudo-scalar field for a simple consideration as it will not affect the finite temperature behavior discussed in\cite{Cui:2013zha}. 

Let us first check the boundary behavior of the solution. Near the UV boundary, one can extract the asymptotic solutions for above four equations Eq.(\ref{eq:eomV}-\ref{eq:eomPST}). For convenience, we replace the radial coordinate $z$ by the dimensionless variable $u$ with $u=z/z_h$.  The two linear independent solutions are found to be:

\begin{eqnarray}\label{asym solu}
\textrm{V} &:&
V_1\to uY_1\left(uz_h\omega\right),\quad V_2\to uJ_1\left(uz_h\omega\right)\\
\textrm{AV} &:&
A_1\to uY_1\left(uz_h\sqrt{\omega^2-4A^2\pi^2}\right),\quad A_2\to uJ_1\left(uz_h\sqrt{\omega^2-4A^2\pi^2}\right)\\
\textrm{S} &:&
S _1\to u^2
   J_1\left(uz_h\sqrt{2\mu_g^2+\omega^2}\right),
\quad
S _2\to u^2
  Y_1\left(uz_h\sqrt{2\mu_g^2+\omega^2}\right)
 \\
\textrm{PS} &:&
\pi _1\to u
   J_1\left(\frac{u \omega }{z_{h}}\right),
\quad
\pi _2\to u
   Y_1\left(\frac{u \omega }{z_{h}}\right)
\end{eqnarray}
Here $J_1$ and $Y_1$ are the first-kind Bessel function and second-kind Bessel function respectively. 
As discussed in \cite{retarded green}, in the Minkowski space-time, the choice of in-falling boundary condition at the horizon selects the retarded Green's function:
\begin{eqnarray}\label{infalling}
K_{-}\to(1-u)^{-i\frac{z_{h}\omega}{4}}
\end{eqnarray}

The solutions of equations of motion can be expressed by the combination of the two independent asymptotic solutions: $K_1 = (V_1,\; A_1\; S_1,\; \pi_1)$ and $K_2 = (V_2,\; A_2\; S_2,\; \pi_2)$
\begin{eqnarray}
\label{solution}
K(u)=A(\omega,q)K_{1}(\omega,q,u)+B(\omega,q)K_{2}(\omega,q,u)\longrightarrow(1-u)^{-i\frac{z_{h}\omega}{4}}
\end{eqnarray}
where the coefficients $A(\omega,q)$ and $B(\omega,q)$ are fixed by the IR in-falling boundary condition at the horizon.
The retarded Green's function can be obtained from the dual bulk fields. As an illustration, for scalar fields, one writes the on shell action which reduces to surface terms:
\begin{equation}
S=\int\frac{d^{4}p}{\left(2\pi\right)^{4}}\left.e^{-\Phi(z)}f(z)a(z)^{\frac{3}{2}}S(p,z)\partial_{z}S(p,z)\right|_{z=0}^{z=z_{h}},\label{surface}
\end{equation}
Following the prescription in \cite{retarded green}, after substitute Eq.(\ref{solution}) into surface terms of the on shell action, one can find that the spectral function which is related to the imaginary part of two point retarded Green's function is proportional to the imaginary part of $B(\omega,q)/A(\omega,q)$.
\begin{eqnarray}
\rho(\omega,q)=-\frac{1}{\pi}\mathrm{Im}\, G(\omega,q)\,\theta(\omega^{2}-q^{2})\varpropto\mathrm{Im}\,\frac{B(\omega,q)}{A(\omega,q)},
\end{eqnarray}

\begin{figure}[!h]
\begin{center}
\includegraphics[width=64mm,clip]{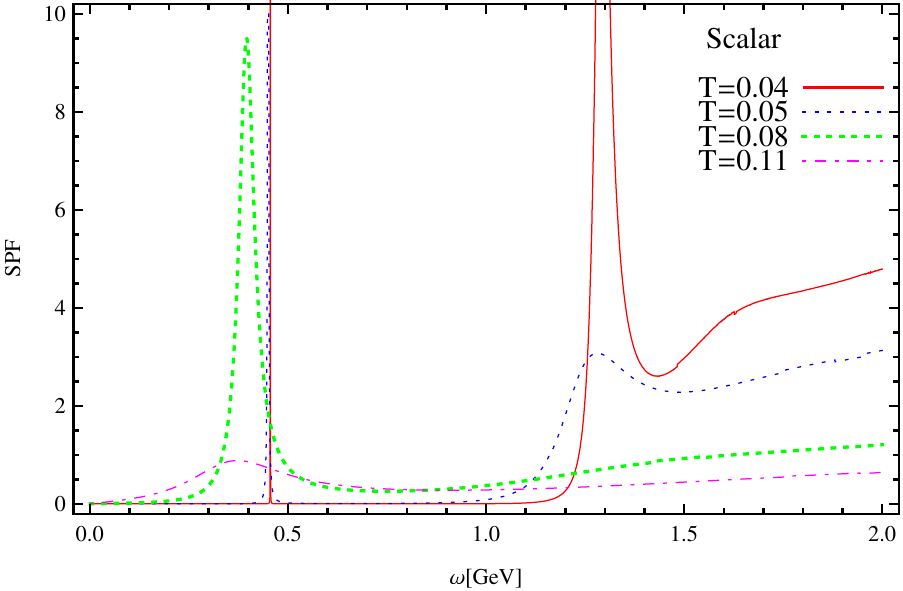}
\includegraphics[width=64mm,clip]{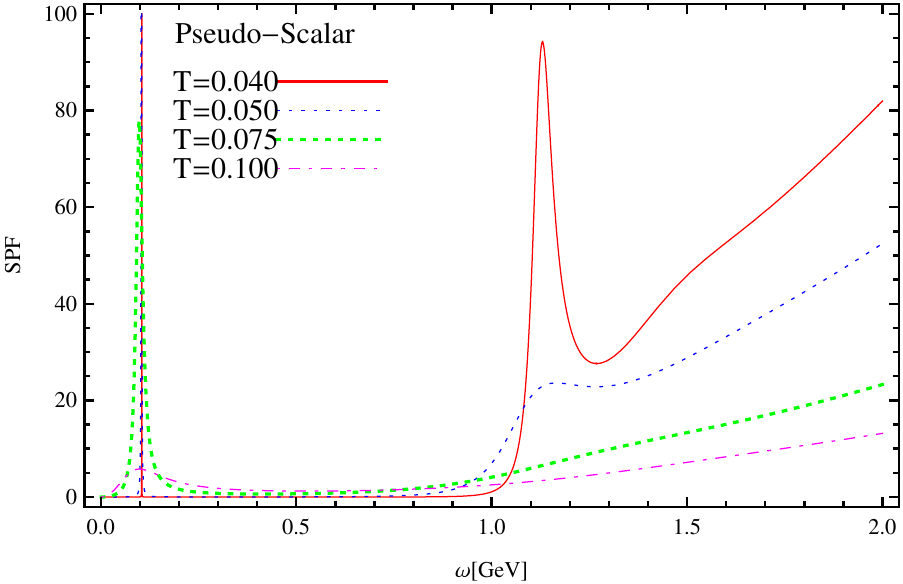}
\includegraphics[width=64mm,clip]{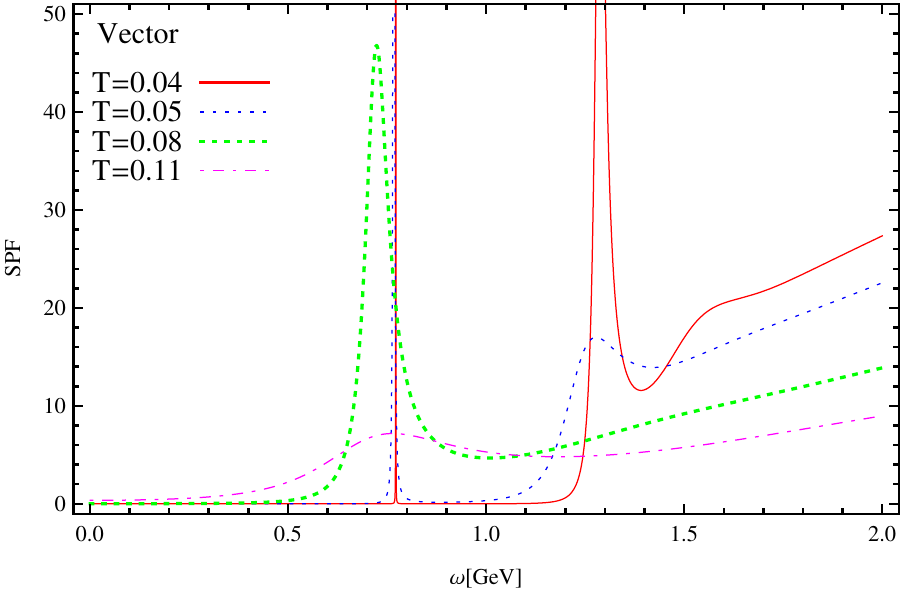}
\includegraphics[width=64mm,clip]{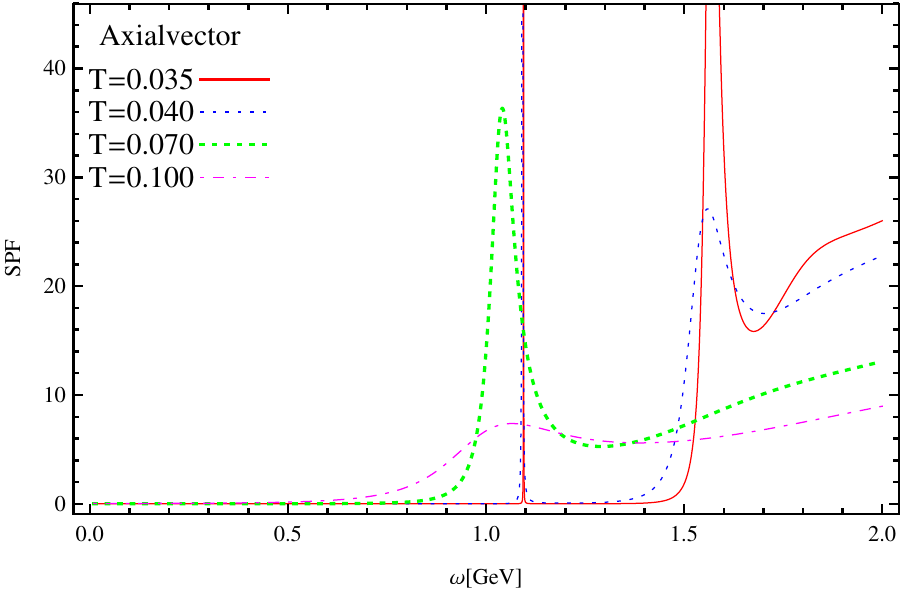}
\end{center}
\caption{The results of spectral function for scalar meson (top, left), pseudo-scalar meson (top,right), vector meson(bottom, left) and axial-vector meson(bottom, right).
}\label{Fig of spectral}
\end{figure}

The numerical results of spectral function for scalar, pseudo-scalar, vector and axial-vector mesons are shown in Fig.\ref{Fig of spectral}.
It can be seen from the results that in low temperature region the peaks which correspond to the poles of the Green's function represent resonance mesons with their masses coinciding to the ones given at zero temperature\cite{Cui:2013xva}. As the temperature increases, the meson states become unstable. It can be seen from the peaks which are shifted towards smaller values and the widths which become broader. Quantitatively, we can get more information by fitting the spectral function with a Breit-Wigner form:
\begin{eqnarray}\label{BWform}
\frac{a \omega^b}{(\omega^2-m^2)^2+\Gamma^2} + P(\omega^2).
\end{eqnarray}
where $m$ and $\Gamma$ are the location and width of the peak respectively. $P(\omega^2)$ is representing a continuum which is taken the form $P(\omega^2)=c_1+c_2\omega^2+c_3(\omega^4)$
The melting temperature or the critical temperature can be defined from the Breit-Wigner form.
That is, if the width of the peak is larger than its height, we can say that no peak can be distinguished anymore. The condition is shown as follow:
\begin{equation}\label{critical temperature}
   h=\frac{a \omega^b}{(\omega^2-m^2)^2+\Gamma^2}\bigg|_{\omega\to m},\qquad  h<\Gamma \; .
\end{equation}
Note that this definition of critical temperature is vague and subjective. In this paper, we will give the range of critical temperature by the condition: $\Gamma/2<h<\Gamma$.
The range of critical temperatures of scalar, pseudo-scalar, vector and axial-vector mesons are shown in Table.\ref{Table:tc}.
\begin{table}[!h]
\begin{center}
\begin{tabular}{|c|c|c|c|c|}
\hline\hline
Meson &Scalar &Pseudo-Scalar & Vector & Axial-Vector   \\
\hline
$T_c$(MeV)&133-136 & 135-140 & 136-140 & 143-146    \\
\hline\hline
\end{tabular}
\caption{The critical temperatures of scalar, pseudo-scalar, vector and axial-vector mesons}
\label{Table:tc}
\end{center}
\end{table}

The results of melting temperature imply that the mesonic quasiparticle state is dissolved around $T_c=140\MeV$ in above considerations. It is noted that the bulk coordinate $z$ plays the role of the running energy scale in boundary theory. As the Hawking temperature increases to around $T_c\simeq140\MeV$, the allowed value for $z$ is given by $0<z<1/(\pi T)\simeq 1/439 \MeV^{-1}$. Such a small value of z will cause the bVEV $v(z)$ with $m_q\simeq 0$ approaches to zero as the power $z^3$ for the condensation $\sigma$. It can be understood that the vanishing bVEV $v(z)$ which corresponds to the chiral condensation plays an important role in the dissolving of mesonic bound state. It can be deduced that these critical behaviors could be the sign of chiral symmetry restoration.

\begin{figure}[!h]
\begin{center}
\includegraphics[width=64mm,clip]{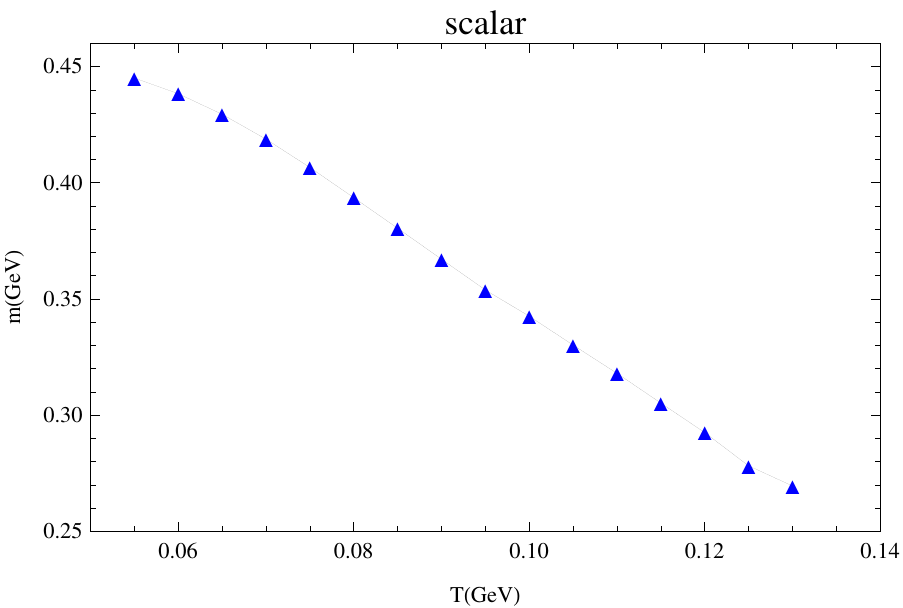}
\includegraphics[width=64mm,clip]{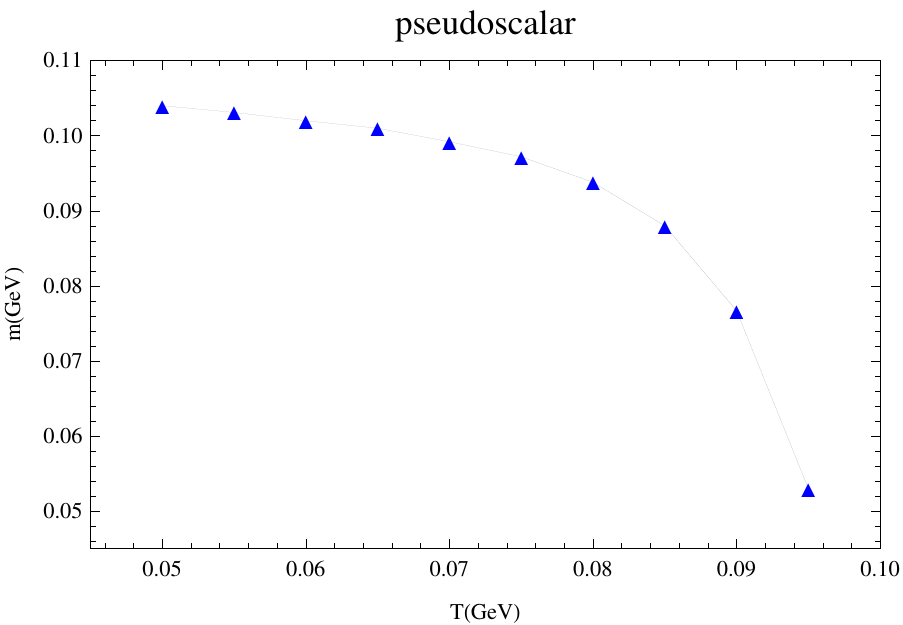}
\includegraphics[width=64mm,clip]{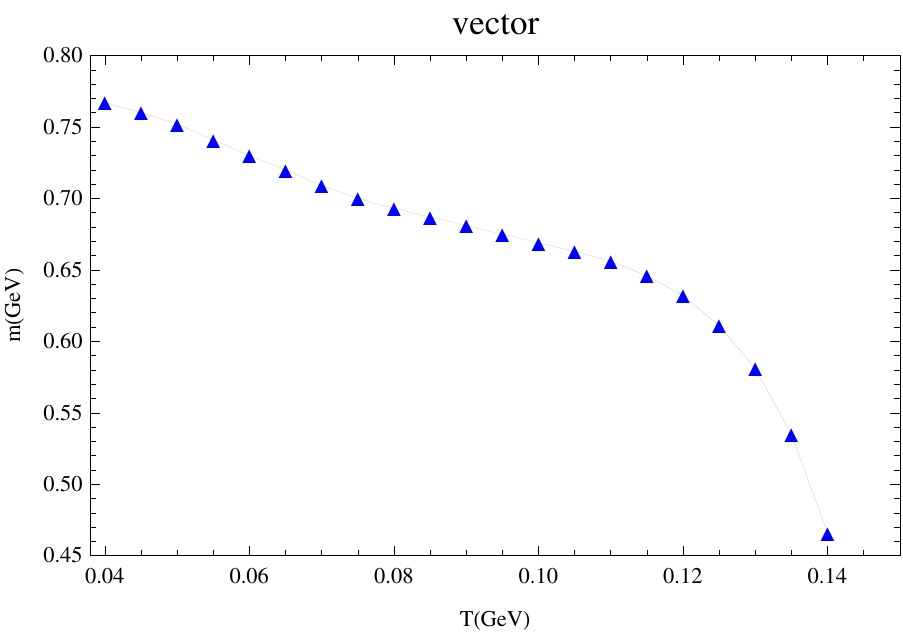}
\includegraphics[width=64mm,clip]{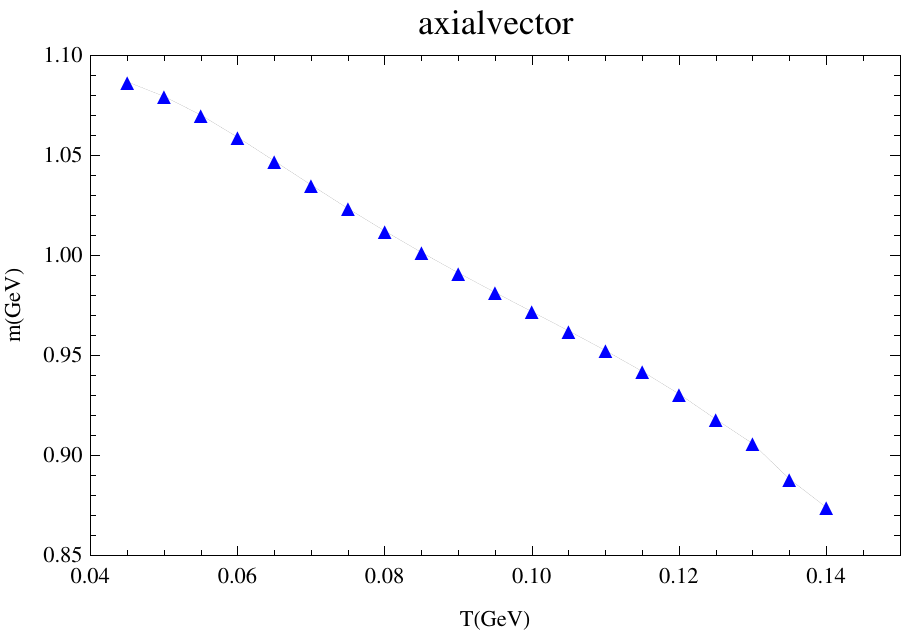}
\end{center}
\caption{The relation between the location of the first peak and the temperature for scalar meson (top, left), pseudo-scalar meson (top,right), vector meson(bottom, left) and axial-vector meson(bottom, right).
}\label{Fig:mass shift}
\end{figure}

From the Breit-Wigner form, we can determine quantitatively the relation between the mass of mesons and the temperature. The results are shown in Fig.\ref{Fig:mass shift}.
It can be seen explicitly that as temperature increases the masses of mesons decrease linearly in low temperature region ($40\sim100\MeV$). Note that around critical temperature $140\MeV$, the spectral function becomes so flat that the numerical fitting has a big ambiguity. It is believed that the mass of scalar and pseudo-scalar mesons will increase slightly around critical temperature, though we can not see here for the large ambiguity. While for vector and axial-vector, the decreasing of mass in medium agrees with other analysis \cite{dsmofv1,dsmofv2,dsmofv3}. The more precise way to study the dependences of temperature is to calculate the quasinormal modes of mesons. We leave it for future study.

\section{back-reaction effects of bulk vacuum}
\label{Chap:back-reaction}

In this section, we will investigate the back-reaction effects of bulk vacuum which includes the quark mass and condensate. In \cite{xie:2006gt}, a fully back-reacted holographic QCD has been constructed. It was found that the back reaction has only small effects on meson spectra.  It is interesting to check its influence on the mass spectra with finite temperature. Let us begin with the following 5-dimensional action,
\begin{equation}
\label{back reaction action}
 S=\int d^5 x \sqrt{\hat{g}} \left( -\hat{R} + {\rm{Tr}}\left[|DX|^{2}+V(X)\right]\right),
\end{equation}
For simplicity we do not take the dilaton field into account in the action. $\hat{R}$ is the five dimensional Ricci scalar. $X$ is the bulk scalar field in Eq.(\ref{action}) with the bulk vacuum expectation form $X=\frac{1}{2}v(z)\textbf{1}_2$. The bVEV $v(z)$ relates to quark mass and condensates in Eq.(\ref{bVEV}) and Eq.(\ref{ABC}). After taking the trace, the action is rewritten as follow:
\begin{equation}
\label{back reaction action}
 S=\int d^5 x \sqrt{\hat{g}} \left( -\hat{R} + \frac{1}{2}\partial_Mv\partial^Mv+V(v)\right),
\end{equation}
with $V(v)=\textrm{Tr}\left[V(X)\right]$. To obtain the black hole solution, we consider the deformed AdSBH background,
\begin{equation} ds^{2}=\frac{e^{2A(z)}}{z^{2}}\left(f(z)dt^{2}-d\vec{x}^{2}-\frac{dz^{2}}{f(z)}\right).\label{metric}
\end{equation}
The equations of motion are
\begin{eqnarray}
  {1\over 2} \hat{g}_{MN} \left(-\hat{R}+ \frac{1}{2}\partial_{P} v \partial^{P}
v + V(v)\right) +\hat{R}_{MN} -\frac{1}{2}\partial_{M} v \partial_{N}
v &=&0 \\
  \frac{\partial V(v)}{\partial v}-\frac{1}{\sqrt{\hat{g}}} \partial_M\left( \sqrt{\hat{g}} \hat{g}^{MN} \partial_Nv\right) &=&0
\end{eqnarray}

The $(t,t)$,$(x_1,x_1)$ and $(z,z)$ components of the gravitational field equations are respectively:
\begin{eqnarray}
     A^{''}+A^{'}\left(\frac{f^{'}}{2f}-\frac{2}{z}\right)+A^{'2}+\frac{2}{z^2}+\frac{v^{'2}}{12}-\frac{f^{'}}{2zf}
    -\frac{e^{2A}V(v)}{6z^2f} &=& 0 \label{eq1}\\
  f^{''}+f^{'}\left(6A^{'}-\frac{6}{z}\right)+f\left(6A^{''}+6A^{'2}+\frac{1}{2}v^{'2}+\frac{12}{z^2}-
   \frac{12A^{'}}{z}\right) -\frac{e^{2A}V(v)}{z^2}&=&0 \label{eq2}\\
  A^{'2}+A^{'}\left(\frac{f^{'}}{4f}-\frac{2}{z}\right)+\left(\frac{1}{z^2}-\frac{e^{2A}V(v)+3zf^{'}}
  {12z^2f}-\frac{v^{'2}}{24}\right) &=&0 \label{eq3}
\end{eqnarray}
From Eq.(\ref{eq2}) and Eq.(\ref{eq3}), we can obtain the equation of the warped factor
\begin{equation}\label{warp factor}
    A^{''}-A^{'2}+\frac{2}{z}A^{'}+\frac{1}{6}v^{'2}=0
\end{equation}
This equation cannot be analytically solved with the bVEV $v(z)$ given in Eq.(\ref{bVEV}). We then numerically solve $A(z)$ by using the UV boundary condition $A(0)=0$ and its derivative vanishes for a general situation.

\begin{figure}[!h]
\begin{center}
\includegraphics[width=64mm,clip]{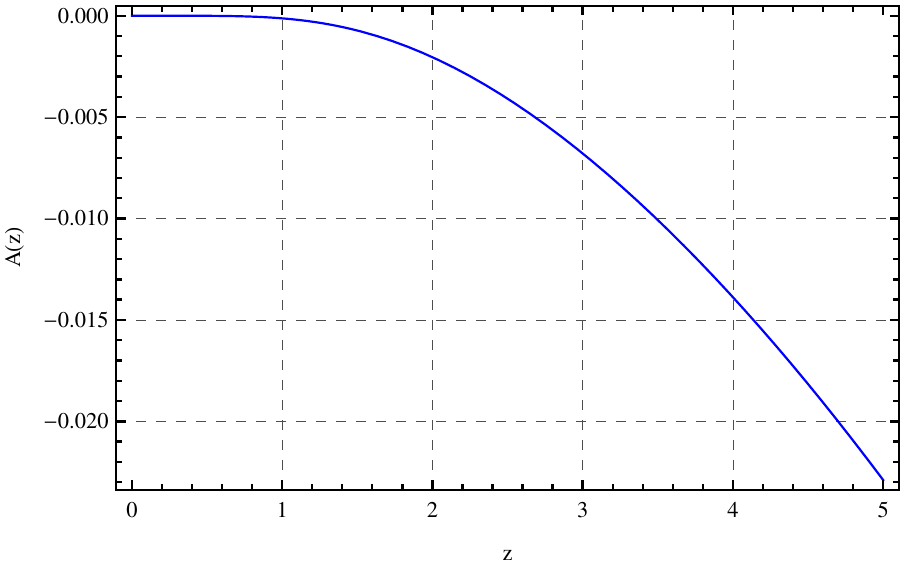}
\includegraphics[width=64mm,clip]{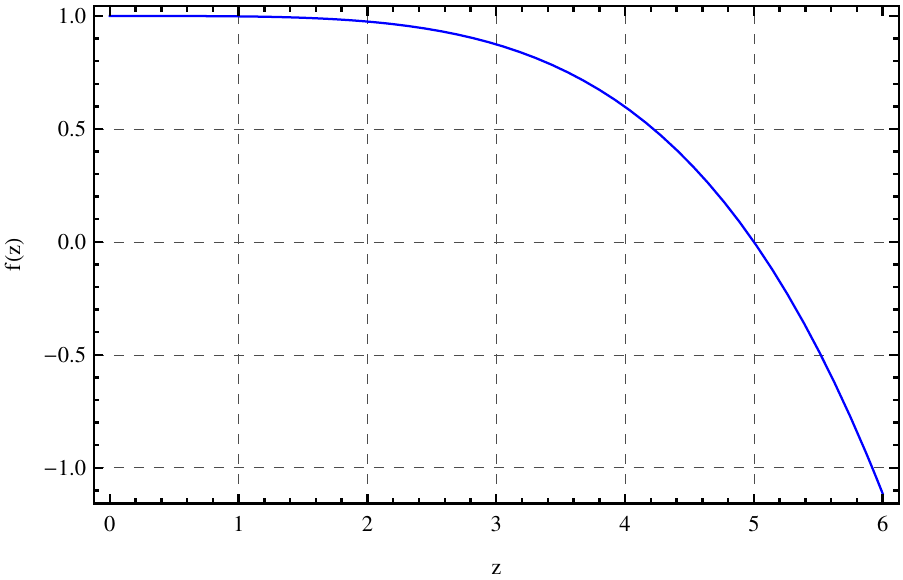}
\end{center}
\caption{The numerical solution for $A(z)$ (left side) and $f(z)$ with horizon $z_h$=5 (right side).
}\label{Fig Az}
\end{figure}

\begin{figure}[!h]
\begin{center}
\includegraphics[width=70mm,clip]{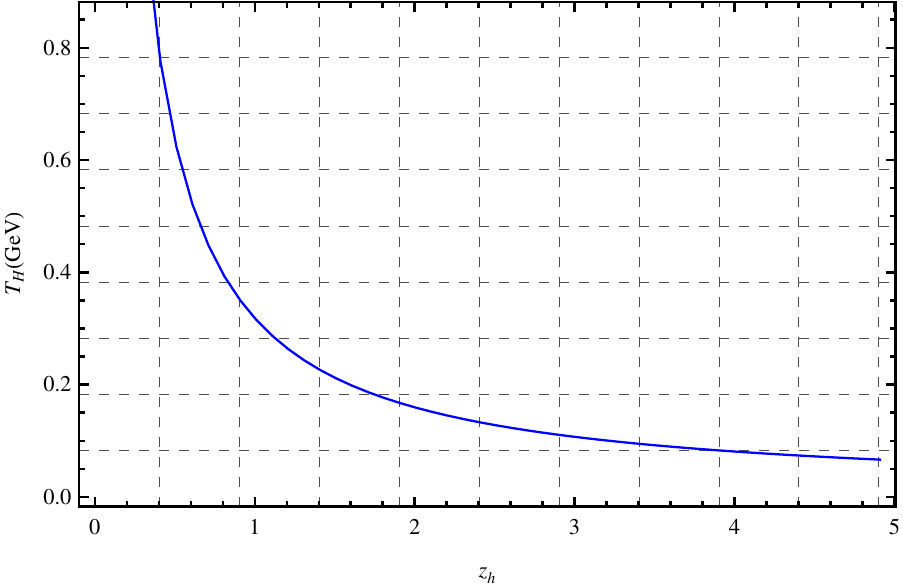}
\end{center}
\caption{The relation between temperature $T_H$ and horizon $z_h$}
\label{Fig Th}
\end{figure}

While from Eq.(\ref{eq1}) and Eq.(\ref{eq2}), one can analytically solve $f(z)$ as:
\begin{equation}\label{f}
    f(z)=C_1+C_2\int_0^ze^{-3A(z)}z^3dz
\end{equation}
where $C_1$ and $C_2$ are integral constants.
Near the boundary $z\rightarrow0$, we require the metric to be asymptotic to $AdS_5$:
\begin{equation}\label{BC0}
    f(0)=1
\end{equation}
Near the horizon $z=z_h$, we require
\begin{equation}\label{BC1}
    f(z_h)=0
\end{equation}
Solution of $f(z)$ can be expressed as
\begin{equation}\label{solution of f}
    f(z)=1-\frac{\int^z_0 x^3 e^{-3A(x)}dx}{\int^{z_h}_0 x^3 e^{-3A(x)}dx}
\end{equation}
One can expand $f(z)$ at the UV boundary with requiring $A(0)=0$,
\begin{equation}
    f(z\to 0)=1-\frac{z^4}{4 \int_0^{z_h} e^{-3A(t)}t^3dt}+\cdots
\end{equation}
Comparing with AdS black-hole solution, it can be seen that the correction of back-reaction contributes to the higher order terms of $f(z)$.
The numerical results of $A(z)$ and $f(z)$ are presented in Fig.\ref{Fig Az}.
It is easy to obtain the Hawking temperature,
\begin{equation}\label{hawking T}
    T_H=-\frac{1}{4\pi}\frac{\partial f}{\partial z}\bigg|_{z\to z_h}=\frac{z_h^3 e^{-3 A(z_h)}}{4
    \pi \int_0^{z_h} e^{-3 A(x)} x^3 dx}
\end{equation}
We plot the temperature $T_H$ v.s. horizon $z_h$ in Fig.\ref{Fig Th}. The monotonous behavior indicates that such a black hole solution is stable. 

With the above analysis, we are now in the position to investigate the finite temperature behavior of mesons after considering the back-reaction effects of bulk vacuum. The action has the same form as Eq.(\ref{action}) except for the background metric, which has been replaced by the back-reaction improved one $\hat{g}$:
\begin{equation}
S=\int d^{5}x\,\sqrt{\hat{g}}e^{-\Phi(z)}\,{\rm {Tr}}\left[|DX|^{2}-m_{X}^{2} |X|^{2}-\lambda_X |X|^{4}-\frac{1}{4g_{5}^2}(F_{L}^2+F_{R}^2)\right],\label{action41}
\end{equation}

\begin{figure}[!h]
\begin{center}
\includegraphics[width=64mm,clip]{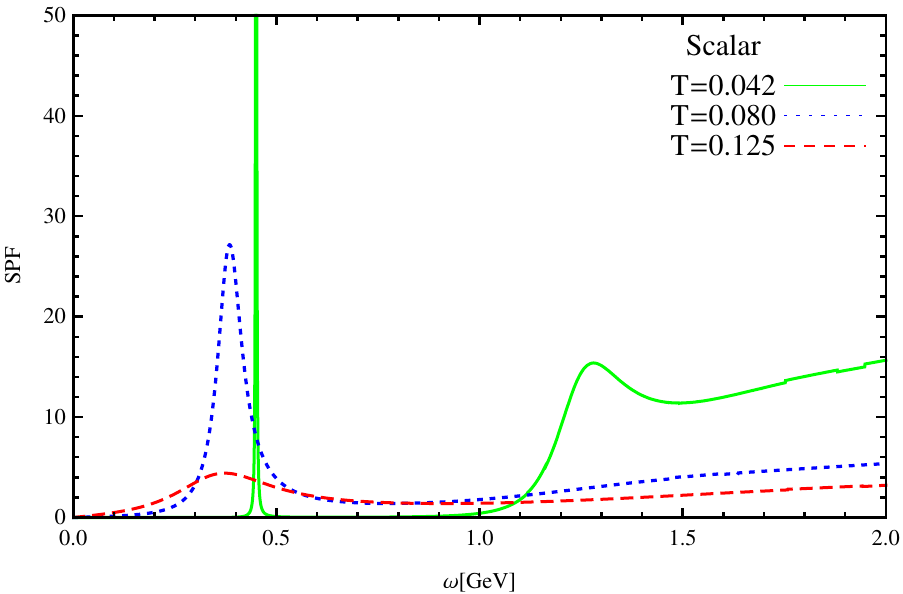}
\includegraphics[width=64mm,clip]{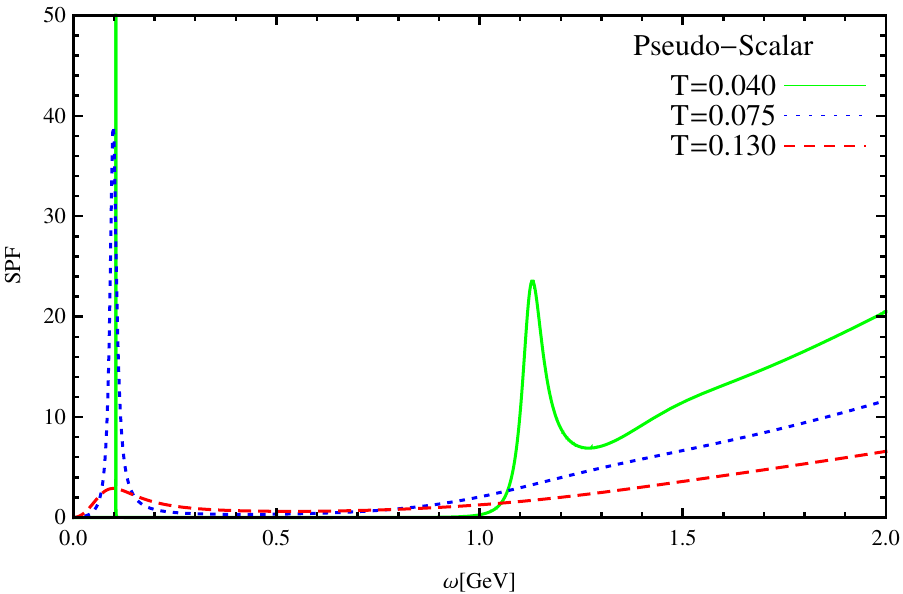}
\includegraphics[width=64mm,clip]{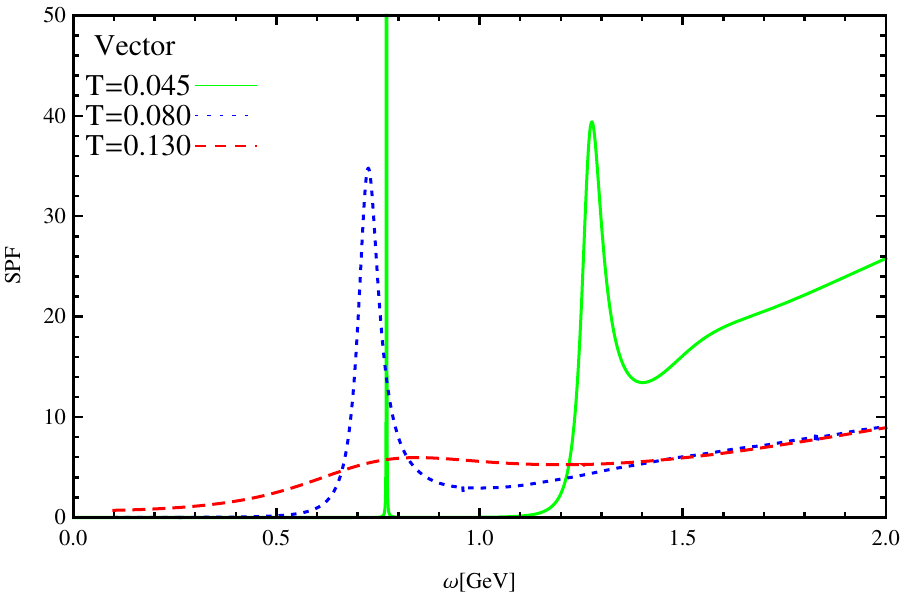}
\includegraphics[width=64mm,clip]{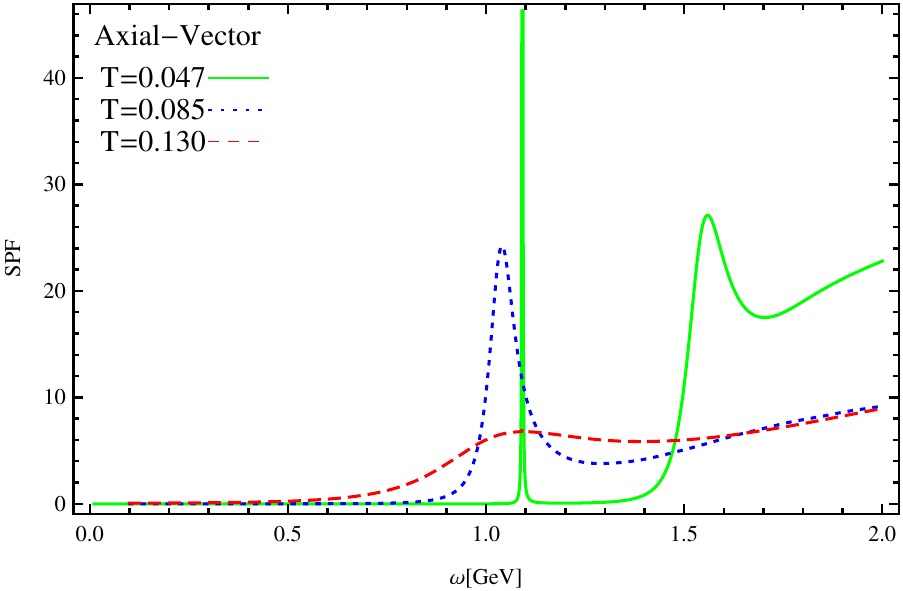}
\end{center}
\caption{The results of spectral function for scalar meson (top, left), pseudo-scalar meson (top,right), vector meson(bottom, left) and axial-vector meson(bottom, right) with back-reaction effects.
}\label{mass spectral br}
\end{figure}

Making a similar calculation as the one in section \ref{Chap:thermal spectral}, we can obtain the mesons' thermal spectral function with back-reaction improved gravity background. The numerical results are shown in Fig.\ref{mass spectral br}. It can be seen that in low temperature region the locations of the peaks are nearly the same as the ones without back-reaction effects in section \ref{Chap:thermal spectral}. Such phenomena agree well with the conclusion in \cite{xie:2006gt}. It is found that the warped factor $A(z)$ shown in Fig.\ref{Fig Az} can well be fitted by a simple form $A(z)=-k^2 z^2$ with $k$ around $k\simeq 30\MeV$.

It is noticed that in zero temperature region $f(z)=1$ and the back-reaction correction of quark mass and condensate provides
very little effects on mass spectra. While in high temperature region, it is seen that the melting temperatures have increased about $10\MeV$. By fitting the spectral function with the Breit-Wigner form in Eq.(\ref{BWform}), we can obtain the critical temperature with including the back-reaction effects of bulk vacuum. The results are presented in Table.\ref{Table:tc with br}

\begin{table}[!h]
\begin{center}
\begin{tabular}{|c|c|c|c|c|}
\hline\hline
Meson &Scalar &Pseudo-Scalar & Vector & Axial-Vector   \\
\hline
$T_c$(MeV)& 142-147 & 143-148 & 148-152 & 151-157    \\
\hline\hline
\end{tabular}
\caption{The critical temperatures of scalar, pseudo-scalar, vector and axial-vector mesons with back-reaction effects}
\label{Table:tc with br}
\end{center}
\end{table}

It should be pointed out that in the above calculation the dilaton field in the action Eq.(\ref{back reaction action}) is still taken as a background field. The back reaction effects of bulk vacuum which includes quark mass and quark condensate have increased the melting temperature to be around 
\begin{equation}
T_c \simeq 150\pm 7 \MeV
\end{equation}
Such a result is consistent with the ones yielded from lattice QCD simulations. In \cite{lattice1}, the chiral and deconfinement critical temperatures were found to be $147\MeV \sim 157\MeV$. In \cite{lattice2}, the chiral transition temperature of two massless flavors was shown to be $T_c = 154\pm9\MeV$. For physics masses of three flavor quarks, the chiral transition temperature was found to be $T_c = 155(1)(8)$ MeV\cite{lattice3}.

\section{Conclusions and Remarks}
\label{Chap:Sum}

 We have investigated the finite temperature behavior of IR-improved bulk holographic AdS/QCD model built recently in\cite{Cui:2013xva}. The spectral function of mesons has been analyzed following the prescription in \cite{retarded green}. By fitting the spectral function with a Breit-Wigner form, the critical temperature of mesons is found to be around $140\MeV$.  It has been noticed that in low temperature region, the peaks which correspond to the poles of the Green's function are consistent with the masses calculated in zero temperature case \cite{Cui:2013xva}. We would like to point out that there exists the vagueness of the critical temperature criterion. In obtaining the critical temperature,  we have to take a range of the melting temperature with the condition between the hight ($h$) and width ($\Gamma$) of peak that: $\Gamma/2<h<\Gamma$. In this paper, we have considered the back-reaction effects of bulk vacuum and yielded an improved metric of background gravity.  The mesons' thermal mass spectral function has been calculated based on the back-reaction improved action, which can lead the critical temperature to be increased about $10$ MeV. A reasonable melting temperature has been found to be $T_c \simeq 150\pm 7$ MeV, which is consistent with the recent results obtained from lattice QCD simulations.

\section*{Acknowledgements}
This work is supported in part by the National Nature Science Foundation of China (NSFC) under Grants No.10975170, No.10905084, No.10821504; and the Project of Knowledge Innovation Program (PKIP) of the Chinese Academy of Science. \\

\appendix

\end{document}